\documentclass[preprint,12pt]{elsarticle}

\usepackage{graphics}

\usepackage{amssymb}

\journal{Annals of Physics}

\begin{document}

\begin{frontmatter}



\title{Exact decoherence dynamics of a single-mode optical field}


\author[NUS,LZU]{Jun-Hong An}
\ead{phyaj@nus.edu.sg}
\address[NUS]{Centre for Quantum Technologies and Department of Physics, National University of Singapore, 3 Science Drive 2, Singapore 117543, Singapore}
\address[LZU]{Department of Modern Physics, Lanzhou University, Lanzhou 730000,
P. R. China} \author[NUS]{Ye Yeo}
\author[NUS]{C. H. Oh}\ead{phyohch@nus.edu.sg}

\begin{abstract}
We apply the influence-functional method of Feynman and Vernon to
the study of a single-mode optical field that interacts with an
environment at zero temperature. Using the coherent-state formalism
of the path integral, we derive a generalized master equation for
the single-mode optical field.  Our analysis explicitly shows how
non-Markovian effects manifest in the exact decoherence dynamics for
different environmental correlation time scales. Remarkably, when
these are equal to or greater than the time scale for significant
change in the system, the interplay between the backaction-induced
coherent oscillation and the dissipative effect of the environment
causes the non-Markovian effect to have a significant impact not
only on the short-time behavior but also on the long-time
steady-state behavior of the system.

\end{abstract}

\begin{keyword}
Non-Markovian decoherence \sep Master equation \sep Influence
functional theory

\PACS 03.65.Yz \sep 03.67.-a


\end{keyword}

\end{frontmatter}


\section{Introduction}
In realistic conditions, it is impossible to completely isolate a
quantum system $S$ from its environment $E$.  A proper analysis of
the quantum dynamics of $S$ must therefore take into account the
decoherence effect of $S$ induced by $E$.  Conventional approaches
to a theoretical study of the dynamics of such an open quantum
system have treated the interactions between $S$ and $E$
perturbatively. Invoking the Born-Markov approximation, these yield
approximate equations of motion such as the Redfield or master
equations \cite{Redfield65, Lindblad76, Carmichael93, Breuer02} for
the reduced system we are interested in.  The Born-Markovian
approximation works well when the environment correlation time
$\tau_E$ is small compared to the time scale $\tau_0$ for
significant change in $S$ \cite{Carmichael93, Breuer02}.  However,
in the light of recent experiments (see, for instance, Refs.
\cite{Dubin07, Koppens07, Mogilevtsev08}), it is evident that there
are many physically relevant situations where the Markovian
assumption does not hold, and a non-Markovian treatment of the open
system dynamics is necessary.  Furthermore, there is a general
interest in the fundamental theory of open quantum systems to extend
the well-developed concepts and methods for Markovian dynamics to
the non-Markovian case.  The development of a general description to
open quantum system has thus attracted much attention lately.  Some
recent works exploring the non-Markovian decoherence properties of
quantum systems include Refs. \cite{Hu92, Maniscalco04, Paris07,
Hu08, Ban06, Goan07, An07, Bellomo07, Breuer06}.

In this paper, we consider the exact non-Markovian decoherence
dynamics of a single-mode optical field system - a basic building
block in quantum communication and computation \cite{Nielsen00}.
Indeed, many of the quantum information processing protocols
\cite{Sch-kitten, Sch-cat}, especially in the field of
continuous-variable quantum information processing
\cite{Braunstein05}, involve optical fields.  In practice, the
optical field inevitably interacts with surrounding environment,
which always results in the decoherence of the optical field.
Actually, this decoherence effect still exists even when the optical
field is transmitted in an optical fiber \cite{Pellizzari97,
Lidar06}. Decoherence will undoubtedly have a detrimental influence
on the performance of the protocols.  It is thus very important for
one to have a complete quantitative understanding of the destructive
influences of the environment. Many of the current quantum optical
experiments are performed at low temperatures, under vacuum
condition. In this case, the main source of decoherence is the
vacuum fluctuation. Many of the theoretical studies on the
decoherence dynamics of an optical field to date rely on the
Born-Markovian approximation \cite{Paris03, Jakub04, An05, Rossi06}.
We note that, only very recently, some phenomenological models on
non-Markovian decoherence dynamics of optical fields have been
studied \cite{Ban06, Goan07, An07}.  Based on perturbation, these
may not capture all the characteristics of the exact non-Markovian
dynamics.  A more satisfactory theory, derived from first
principles, that describes an open optical field system is thus
desirable. To this end, we apply the influence-functional method of
Feynman and Vernon \cite{Feynman63, Caldeira83} to the study of the
system $S$ of a single-mode optical field that interacts with an
environment $E$, consisting of a set of harmonic oscillators at zero
temperature.

The Feynman-Vernon influence functional theory allows one to derive
the exact non-Markovian dynamics of $S$.  In this method, the
density operator of the combined system, $S$ and $E$, is expressed
as a double path integral.  The exact dynamics of the reduced system
$S$ is then obtained by integrating this path integral over the
degrees of freedom of $E$.  The effective action that governs the
evolution of $S$ thus consists of the free action of $S$ and an
influence functional.  All the environmental effects on $S$ are
dynamically incorporated in the influence functional, and both the
backactions from $E$ to $S$ and from $S$ to $E$ can be treated
self-consistently. In our analysis, we mainly address the issues
about how the non-Markovian effect takes action and what its
influence on the decoherence dynamics of the optical field system in
different parameter regimes is.  As an explicit example, we will
study the time evolution of the so-called Schr\"{o}dinger cat state
\cite{Sch-kitten, Sch-cat} and see what an exact treatment of its
decoherence dynamics would yield.

Indeed, it had been shown that non-Markovian effect shows its
significant consequence on the decoherence dynamics just by a
transient oscillation in short time scale \cite{Paris07, Ban06,
Goan07, An07}.  Here, we will show, besides this transient
oscillation in short time scale, the non-Markovian effect can also
influence the behavior of steady state in long-time scale when
$\tau_E \simeq \tau_0$ or $\tau_E > \tau_0$.

Our paper is organized as follows.  In Sec. II, we introduce a model
of the single-mode optical field in an environment, and the
coherent-state representation.  In Sec. III, we present a detailed
derivation of the quantum non-Markovian master equation using
influence functional theory in the coherent-state representation.
Our generalized master equation has time-dependent frequency shift
and decay rate, and it reduces to the general Markovian one under
certain approximation.  Sec. IV is devoted to a numerical study of
the system decoherence dynamics for different environmental
correlation time scales.  In particular, the decoherence dynamics of
the Schr\"{o}dinger cat state \cite{Sch-kitten, Sch-cat} is
investigated explicitly.  Finally, a summary is given in Sec. V.

\section{A single-mode optical field in a quantized radiation field}
We consider a single-mode optical field $S$ that interacted with an
environment $E$. The environment, as usual, is modeled by a
collection of harmonic oscillators. The total Hamiltonian governing
the coupled $S$ and $E$ is given by \cite{Carmichael93}
\begin{equation}\label{HM}
H = H_S + H_E + H_I,
\end{equation}
where
\begin{equation}
H_{S} = \hbar \omega _{0}a^{\dag }a
\end{equation}
is the Hamiltonian of the free single-mode optical field $S$,
\begin{equation}
H_{E} = \hbar \sum_{k}\omega_{k}b_{k}^{\dagger }b_{k}
\end{equation}
is the Hamiltonian of the environment $E$, and
\begin{equation}
H_{I} = \hbar \sum_{k}(g_{k}a^{\dag }b_{k} + g_{k}^{\ast
}ab_{k}^{\dag })
\end{equation}
is a model for the system-environment interaction. $a$ and $a^{\dag
}$ are the annihilation and creation operators of the single-mode
optical field $S$ with frequency $\omega _0$.  Similarly, $b_k$ and
$b_k^{\dag}$ $(k = 1, 2, \cdots)$ are the annihilation and creation
operators of the $k$-th mode of the environment with frequency
$\omega_k$.  The coupling strength between $S$ and the $k$th mode of
$E$ is given by $g_k$. The environment is assumed to be at zero
temperature initially. By this model we will address the decoherence
mechanism of $S$ due to its energy exchange with the modes of the
environmental vacuum fluctuation. We emphasize the model and
formulation developed in the following are also applicable to many
similar systems in quantum optics, for example, an optical field in
a leaky cavity \cite{Carmichael93} or in an optical fiber
\cite{Pellizzari97, Lidar06}, and the noise effect on a
nanomechanical oscillator \cite{Rae08}.

The Hamiltonian $H$ can be expressed, in the coherent-state
representation \cite{Zhang90}, in terms of
\begin{equation}\label{HC}
H(\bar{\alpha},\alpha ,\mathbf{\bar{z}},\mathbf{z}) = \hbar \{\omega
_{0}\bar{\alpha}\alpha +\sum_{k}[\omega
_{k}\bar{z}_{k}z_{k}+(g_{k}\bar{\alpha}z_{k}+g_{k}^{\ast
}\bar{z}_{k}\alpha )]\},
\end{equation}
where $\mathbf{z}$ denotes $(z_{1},z_{2},\cdots )$.  The coherent
states $|\alpha\rangle \equiv \exp(\alpha a^{\dagger})|0\rangle$ and
$|z_k\rangle \equiv \exp (z_{k}b_{k}^{\dagger})|0_k\rangle$ are the
eigenstates of $a$ and $b_k$ respectively: $a|\alpha \rangle =
\alpha |\alpha \rangle$; $b_{k}|z_{k}\rangle = z_{k}|z_{k}\rangle$.
They are nonorthogonal: $\langle\bar{\alpha}|\alpha^{\prime }\rangle
= \exp(\bar{\alpha}\alpha^{\prime})$;
$\langle\bar{z}_{k}|z_{k}^{\prime}\rangle =
\exp(\bar{z}_{k}z_{k}^{\prime})$.  And, they form overcomplete sets:
$\int d\mu \left( \alpha\right) |\alpha \rangle \langle
\bar{\alpha}| = 1 = \int d\mu(z_{k})|z_{k}\rangle \langle
\bar{z}_{\mathbf{k}}|$, with the measures $d\mu\left(\alpha\right) =
\frac{\exp(-\bar{\alpha}\alpha)}{\pi}d^{2}\alpha$ and $d\mu(z_{k}) =
\frac{\exp\left(-\bar{z}_{k}z_{k}\right)}{\pi}d^{2}z_{k}$.  In the
path integral approach, one needs to choose a convenient
representation.  It turns out that employing the coherent-state
representation allows very straightforward evaluation of the path
integrals.  In the following section, we will use this method to
derive the exact decoherence dynamics of the optical field system.

\section{Quantum non-Markovian master equation}
\subsection{The influence functional in coherent-state representation}
Suppose $\rho_T(t)$ describes the state of our single-mode optical
field system $S$ plus the environment $E$ as a whole.  This total
density matrix obeys the Liouville-von Neumann equation
$i\hbar\partial\rho_T(t)/\partial t = [H, \rho_T(t)]$, which gives
the formal solution:
\begin{equation}\label{fs}
\rho_T(t) =
\exp(-\frac{i}{\hbar}Ht)\rho_T(0)\exp(\frac{i}{\hbar}Ht).
\end{equation}
In the coherent-state representation, $\rho_T(t)$ can be expressed
in terms of
\begin{eqnarray}\label{tot}
& &\langle \bar{\alpha}_{f},\mathbf{\bar{z}}_{f}|\rho _{T}\left(
t\right)|\alpha _{f}^{\prime },\mathbf{z}^{\prime}_{f}\rangle
 = \int d\mu(\mathbf{z}_{i})d\mu(\alpha _{i}) d\mu(\mathbf{z}_{i}^{\prime}) d\mu (\alpha _{i}^{\prime}) \nonumber \\
& & \times \langle \bar{\alpha}_{f},\mathbf{\bar{z}}_{f};t|\alpha
_{i},\mathbf{z}_{i};0\rangle \langle
\bar{\alpha}_{i},\mathbf{\bar{z}}_{i}|\rho_{T}(0)|\alpha
_{i}^{\prime },\mathbf{z}_{i}^{\prime }\rangle
\langle \bar{\alpha}_{i}^{\prime},\mathbf{\bar{z}}_{i}^{\prime};0|\alpha _{f}^{\prime},\mathbf{z}^{\prime}_{f};t\rangle,\nonumber  \\
\end{eqnarray}
where the resolution of identity has been used.  Since we are only
interested in the dynamics of $S$, it suffices to work with the
reduced density matrix, which is obtained by integrating over the
environmental variables.  This can be expressed in terms of
\begin{eqnarray}\label{rout}
&&\rho(\bar{\alpha}_{f},\alpha _{f}^{\prime };t)
 =  \int d\mu (\mathbf{z}_{f})\langle \bar{\alpha}_{f},\mathbf{\bar{z}}_{f}|\rho _{T}\left(t\right) |\alpha _{f}^{\prime },\mathbf{z}_{f}\rangle
\nonumber \\
& = & \int d\mu (\alpha _{i})d\mu(\alpha _{i}^{\prime
})\mathcal{J}(\bar{\alpha}_{f},\alpha _{f}^{\prime
};t|\bar{\alpha}_{i},\alpha _{i}^{\prime };0) \rho
(\bar{\alpha}_{i},\alpha_{i}^{\prime };0).
\end{eqnarray}
Here, we have assumed that the initial total density matrix factors
into a system part and an environment part, i.e., $\rho_T(0) =
\rho(0) \otimes \rho_E(0)$. Now, it remains to determine the
effective propagating function for the reduced density matrix,
\begin{eqnarray}\label{effp}
&&\mathcal{J}(\bar{\alpha}_{f}, \alpha _{f}^{\prime };
t|\bar{\alpha}_{i}, \alpha_{i}^{\prime}; 0)
 =  \int d\mu(\mathbf{z}_{f})d\mu(\mathbf{z}_{i})d\mu(\mathbf{z}_{i}^{\prime}) \nonumber \\
& & \times \langle\bar{\alpha}_{f}, \mathbf{\bar{z}}_{f};
t|\alpha_{i}, \mathbf{z}_{i}; 0\rangle
\rho_{E}(\mathbf{\bar{z}}_{i},\mathbf{z}_{i}^{\prime}; 0)
\langle\bar{\alpha}_{i}^{\prime},
\mathbf{\bar{z}}_{i}^{\prime};0|\alpha_{f}^{\prime},\mathbf{z}_{f};t\rangle.
\end{eqnarray}
Equation (\ref{effp}) contains the forward and backward propagators
of the total system.  These can be expressed as path integrals.  To
evaluate the forward propagator $\langle\bar{\alpha}_{f},
\mathbf{\bar{z}}_{f}; t|\alpha_{i}, \mathbf{z}_{i}; 0\rangle$, one
divides the time interval $t_f - t_i$ into $N$ equal subintervals.
This is followed by inserting $N-1$ copies of the resolution of
identity, each between a subinterval, and taking the limit of $N$
large.  The path integral representation of the forward propagator
can then be obtained:
\begin{equation}\label{prop}
\langle \bar{\alpha}_{f},\mathbf{\bar{z}}_{f};t|\alpha
_{i},\mathbf{z}_{i};0\rangle = \int D^{2}\mathbf{z}D^{2}\alpha \exp
(\frac{i}{\hbar }S[\mathbf{\bar{z}},\mathbf{z},\bar{\alpha},\alpha
]),
\end{equation}
with
\begin{equation}\label{actio}
S[\mathbf{\bar{z}},\mathbf{z},\bar{\alpha},\alpha ] =
S_{S}[\bar{\alpha},\alpha ] +
S_{I}[\mathbf{\bar{z}},\mathbf{z},\bar{\alpha},\alpha ] +
S_{E}[\mathbf{\bar{z}},\mathbf{z}],
\end{equation}
where $S_S$, $S_E$, and $S_I$ are the (complex) actions
corresponding to $H_{S}$, $H_{E}$, and $H_{I}$ respectively.  All
the functional integrations are evaluated over paths
$\mathbf{\bar{z}}(\tau )$, $\mathbf{z}(\tau )$, $\bar{\alpha}(\tau
)$, and $\alpha (\tau )$ with endpoints $\mathbf{\bar{z}}(t) =
\mathbf{\bar{z}}_{f}$, $\mathbf{z}(0)=\mathbf{z}_{i}$,
$\bar{\alpha}(t) = \alpha _{f}$, and $\alpha (0)=\alpha _{i}$.  The
backward propagator $\langle\bar{\alpha}_{i}^{\prime},
\mathbf{\bar{z}}_{i}^{\prime};0|\alpha_{f}^{\prime},\mathbf{z}_{f};t\rangle$
can be evaluated in the same fashion. Substituting Eq. (\ref{prop})
and a similar expression for the backward propagator into Eq.
(\ref{effp}) we obtain
\begin{eqnarray}
&&\mathcal{J}(\bar{\alpha}_{f}, \alpha _{f}^{\prime};
t|\bar{\alpha}_{i}, \alpha_{i}^{\prime}; 0) =
\int D^{2}\alpha D^{2}\alpha^{\prime}\ \exp\{\frac{i}{\hbar}(S_{S}[\bar{\alpha}, \alpha]\nonumber \\
&&~~~~~~~~~~- S_{S}^{\ast}[\bar{\alpha}^{\prime},
\alpha^{\prime}])\} \mathcal{F}[\bar{\alpha}, \alpha,
\bar{\alpha}^{\prime}, \alpha^{\prime}],
\end{eqnarray}
where
\begin{eqnarray}\label{influ}
&&\mathcal{F}[\bar{\alpha}, \alpha, \bar{\alpha}^{\prime},
\alpha^{\prime}]
 =  \int d\mu(\mathbf{z}_{f}) d\mu(\mathbf{z}_{i}) d\mu(\mathbf{z}_{i}^{\prime}) D^{2}\mathbf{z} D^{2}\mathbf{z}^{\prime}
\nonumber\\& & ~~ \times \rho_{E}(\mathbf{\bar{z}}_{i},
\mathbf{z}_{i}^{\prime}; 0) \exp\{\frac{i}{\hbar}(
S_{E}[\mathbf{\bar{z}},\mathbf{z}] +
S_{I}[\mathbf{\bar{z}},\mathbf{z}, \bar{\alpha}, \alpha]
\nonumber\\&
&~~~~~~-S_{E}^{\ast}[\mathbf{\bar{z}}^{\prime},\mathbf{z}^{\prime}]
- S_{I}^{\ast}[\mathbf{\bar{z}}^{\prime},\mathbf{z}^{\prime},
\bar{\alpha}^{\prime}, \alpha^{\prime}])\},
\end{eqnarray}
is the influence functional containing all the environmental effects
on $S$.

\subsection{Evaluation of the influence functional and effective propagating function}
Now, we calculate explicitly the influence functional of our model.
Using the Feynman's procedure, one can obtain the path integral
representation of the forward propagator, Eq.(\ref{prop}), for our
system with the component actions as
\begin{eqnarray}\label{act}
& & S_{S}[\bar{\alpha},\alpha ] =
-i\hbar\bar{\alpha}\alpha \left(t\right) +\int_{0}^{t}d\tau \lbrack i\hbar\bar{\alpha}\dot{\alpha}(\tau )-H_{S}(\bar{\alpha},\alpha )]\},  \nonumber \\
& & S_{E}[\mathbf{\bar{z}},\mathbf{z}] =
\sum_{k}\{-i\hbar\bar{z}_{k}z_{k}(t)+\int_{0}^{t}d\tau \lbrack i\hbar\bar{z}_{k}\dot{z}_{k}(\tau )-H_{E}(\mathbf{\bar{z}},\mathbf{z})]\},  \nonumber \\
& & S_{I}[\mathbf{\bar{z}},\mathbf{z},\bar{\alpha},\alpha ] =
-\int_{0}^{t}d\tau H_{I}(\bar{\alpha},\alpha
,\mathbf{\bar{z}},\mathbf{z}).
\end{eqnarray}
The path integral with respect to the environmental variables
$\mathbf{z}$ can be evaluated by the saddle point method under the
boundary conditions $z_{k}(0) = z_{ki}$, $\bar{z}_{k}(t) =
\bar{z}_{kf}$. We have the equations of motion as
\begin{equation}\label{ss}
\dot{z}_k + i\omega_k z_{k} = -ig_k^{\ast}\alpha,
~~\dot{\bar{z}}_{k} - i\omega_{k}\bar{z}_{k} = ig_{k}\bar{\alpha},
\end{equation}
where $\alpha$ and $\bar{\alpha}$ are treated as external sources.
Substituting the solution of Eqs. (\ref{ss}) into Eq. (\ref{prop}),
one can determine the desired path integral.  It is noted that the
prefactor under the contribution of stationary path in the
coherent-state representation is equal to one, and the saddle point
approach to the evaluation of the environmental part here is exact
\cite{Klauder79}.  The path integral with respect to the
environmental variables $\mathbf{z}^{\prime}$, in Eq. (\ref{influ}),
can similarly be obtained.  As explained in our introduction, we
take the environment to be at zero temperature, i.e.
$\rho_{E}(\mathbf{\bar{z}}_i, \mathbf{z}_i^{\prime}; 0) = 1$.
Together with the results of the above path integrals, Eq.
(\ref{influ}) yields the influence functional
\begin{eqnarray}
&&\mathcal{F[}\bar{\alpha},\alpha ,\bar{\alpha}^{\prime },\alpha
^{\prime}] = \exp \{ \int_{0}^{t}d\tau \int_{0}^{\tau }d\tau
^{\prime }[\mu (\tau -\tau^{\prime }) \nonumber(\bar{\alpha}^{\prime
}
\\&&-\bar{\alpha})(\tau )\alpha (\tau ^{\prime}) + \mu ^{\ast }(\tau
-\tau ^{\prime })\bar{\alpha}^{\prime }(\tau ^{\prime})(\alpha
-\alpha ^{\prime })(\tau )]\},
\end{eqnarray}
with $\mu (x) \equiv \sum_{k}e^{-i\omega _{k}x}\left\vert
g_{k}\right\vert ^{2}$.  Here, we have repeatedly used the Gaussian
integral identity $\int \frac{d^{2}z}{\pi }e^{-\gamma
\bar{z}z+\delta z}f(\bar{z})=\frac{1}{\gamma }f(\frac{\delta
}{\gamma })$.

In the derivation of the influence functional above, we have treated
both the backactions of the environment on the system and the system
on the environment self-consistently.  All the effects of the
environment $E$ on $S$ are collected in the influence functional,
which results in a correction to the action of the free single-mode
optical field $S$,
\begin{eqnarray} \label{J}
&&\mathcal{J}(\bar{\alpha}_{f}, \alpha_{f}^{\prime};
t|\bar{\alpha}_{i}, \alpha_{i}^{\prime}; 0)
 =  \int D^{2}\alpha D^{2}\alpha^{\prime}\
\exp\{\bar{\alpha}\alpha(t)\nonumber \\& & ~~~~ +
\bar{\alpha}^{\prime}\alpha ^{\prime}(t) -
\int_{0}^{t}d\tau[\bar{\alpha}\dot{\alpha} +
\dot{\bar{\alpha}}^{\prime}\alpha^{\prime} +
iH_{S}(\bar{\alpha},\alpha)\nonumber \\& &~~~~~~ -
H_{S}(\bar{\alpha}^{\prime},\alpha ^{\prime})]\}\mathcal{F}
[\bar{\alpha}, \alpha , \bar{\alpha}^{\prime}, \alpha^{\prime}].
\end{eqnarray}
To evaluate the path integral in Eq. (\ref{J}), we again employ the
saddle point method and obtain the two equations of motion:
\begin{eqnarray}\label{eom}
0 & = & \dot{\alpha}+i\omega _{0}\alpha +\int_{0}^{\tau }d\tau
^{\prime}\mu\left( \tau -\tau ^{\prime }\right) \alpha \left( \tau
^{\prime }\right) ,
\nonumber \\
0 & = & \dot{\bar{\alpha}}^{\prime }-i\omega
_{0}\bar{\alpha}^{\prime}+\int_{0}^{\tau }d\tau ^{\prime }\mu ^{\ast
}\left( \tau -\tau ^{\prime }\right) \bar{\alpha}^{\prime }\left(
\tau ^{\prime }\right),
\end{eqnarray}
with the boundary conditions $\alpha \left( 0\right) = \alpha _{i}$,
$\bar{\alpha}^{\prime }\left( 0\right) =\bar{\alpha}_{i}^{\prime }$.
The integro-differential equations render the reduced dynamics
non-Markovian, with the memory effect of the environment registered
in the kernel that is nonlocal in time.  The solution of the
integro-differential equations (\ref{eom}) can be expressed in terms
of a complex function $u(\tau)$ as
\begin{equation}\label{soluen}
\alpha(\tau) = \alpha_{i}u(\tau), ~~~~\bar{\alpha}^{\prime}(\tau) =
\bar{\alpha}_{i}^{\prime}\bar{u}(\tau),
\end{equation}
with the boundary condition $u(0) = 1$.  Substituting Eqs.
(\ref{soluen}) into Eq. (\ref{J}) and using Eqs. (\ref{eom}), we
obtain the expression of the effective propagating function of the
reduced system as
\begin{eqnarray}\label{prord}
\mathcal{J}(\bar{\alpha}_{f},\alpha _{f}^{\prime
};t|\bar{\alpha}_{i},\alpha_{i}^{\prime };0) &=& \exp\{
[u\bar{\alpha}_{f}\alpha _{i} + \bar{u}\bar{\alpha}_{i}^{\prime
}\alpha _{f}^{\prime } \nonumber \\ &&+ (1 - \left\vert u\right\vert
^{2})\bar{\alpha}_{i}^{\prime }\alpha _{i}]\},
\end{eqnarray}
where the dependence of $u$ on time is not shown explicitly for
abbreviation.
\begin{figure*}
\begin{center}
\includegraphics[scale = 0.53]{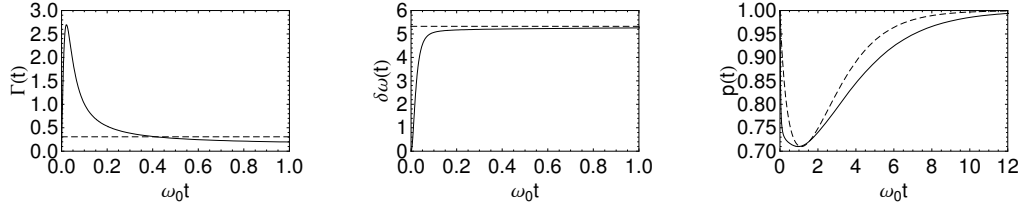}
\end{center}
\caption{\label{fr1} Comparison of the decay rate $\Gamma(t)$ (left)
and the frequency shift $\delta\omega(t)$ (middle), and the purity
(right) of the Schr\"{o}dinger cat state between the non-Markovian
(solid line) and Markovian (dashed line) results in the weak
coupling and short environmental correlation time regime.  The
parameters $\omega_{c}/\omega_{0}=50.0$, $\eta=0.1$, and
$\beta_0=1.0$ are used in the numerical calculation.}
\end{figure*}

\begin{figure*}
\begin{center}
\includegraphics[scale = 0.53]{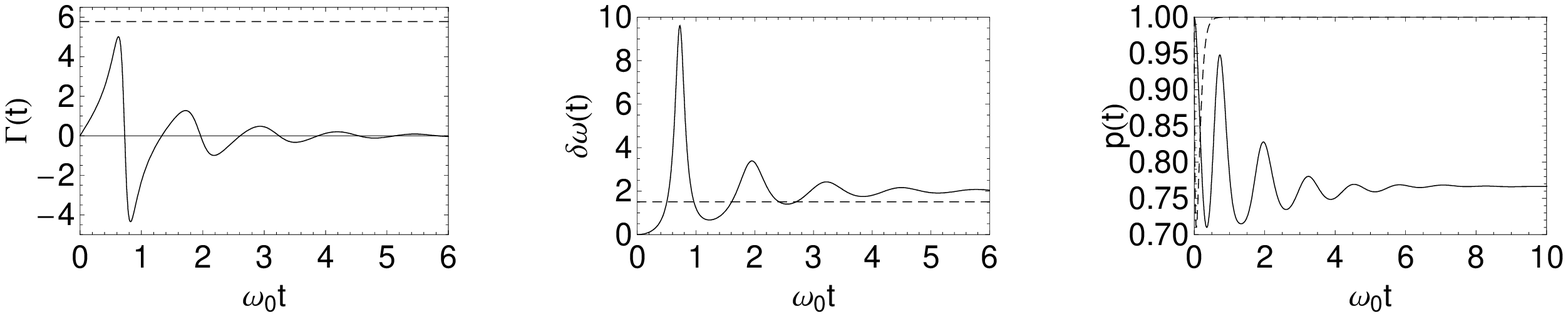}
\end{center}
\caption{\label{fre} Comparison of the decay rate $\Gamma(t)$
(left), frequency shift $\delta\omega(t)$ (middle), and purity
$p(t)$ (right) of the Schr\"{o}dinger cat state between the
non-Markovian (solid line) and Markovian (dashed line) results in
the strong coupling and long environmental correlation time regime.
The parameters $\protect\omega_{c}/\omega_{0}=1.0$, $\eta=5.0$, and
$\beta_0=1.0$ are used in the numerical calculation.}
\end{figure*}

\begin{figure*}
\begin{center}
\includegraphics[scale=0.53]{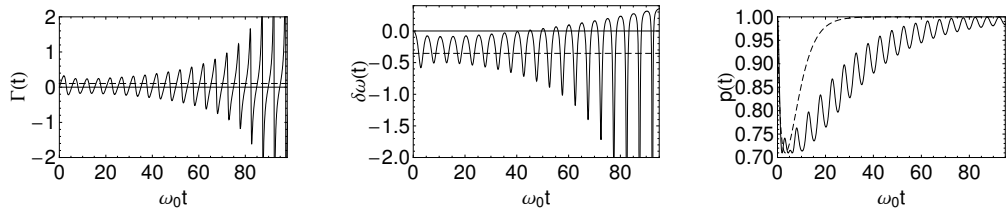}
\end{center}
\caption{\label{fre3} Comparison of the decay rate $\Gamma(t)$
(left), frequency shift $\delta\omega(t)$ (middle), and purity
$p(t)$ (right) of the Schr\"{o}dinger cat state between the
non-Markovian (solid line) and Markovian (dashed line) results when
$\tau_E \gg \tau_0$.  The parameters $\omega_{c}/\omega_{0}=0.2$,
$\eta=5.0$, and $\beta_0=1.0$ are used in the numerical
calculation.}
\end{figure*}
\subsection{The non-Markovian master equation}
Now we can derive the master equation by computing the time
derivative of Eq. (\ref{rout}).  First, from Eq. (\ref{prord}), we
can write down the following identities
\begin{equation}\label{fi}
\alpha _{i}\mathcal{J}=\frac{1}{u}\frac{\delta \mathcal{J}}{\delta
\bar{\alpha}_{f}},~~ \bar{\alpha}_{i}^{\prime
}\mathcal{J}=\frac{1}{\bar{u}}\frac{\delta \mathcal{J}}{\delta
\alpha _{f}^{\prime }},
\end{equation}
which will be used to remove from the time derivative of
$\mathcal{J}$ its dependence on $\alpha_i$ and
$\bar{\alpha}_i^{\prime}$. After taking time derivative to Eq.
(\ref{rout}) and substituting Eqs. (\ref{prord}) and (\ref{fi}) into
it, we obtain the evolution equation
\begin{eqnarray}
&&\dot{\rho}(\bar{\alpha},\alpha ^{\prime};t) =
\{-i\Omega(t)[\bar{\alpha}\frac{\delta\rho(\bar{\alpha}, \alpha
^{\prime};t)}{\delta\bar{\alpha}} -
\frac{\delta\rho(\bar{\alpha}, \alpha ^{\prime};t)}{\delta\alpha}\alpha] \nonumber\\
& & + \Gamma(t)[2\frac{\delta^{2}\rho(\bar{\alpha},
\alpha^{\prime};t)}{\delta\bar{\alpha}\delta\alpha } -
\bar{\alpha}\frac{\delta\rho(\bar{\alpha},\alpha^{\prime};t)}{\delta\bar{\alpha}}
- \frac{\delta\rho(t)}{\delta\alpha}\alpha]\},
\end{eqnarray}
where
\begin{equation}\label{go}
\frac{\dot{u}}{u} \equiv -\Gamma (t)-i\Omega (t),
\end{equation}
and from here on we drop the subscript $f$.

Next, we introduce the following functional differential relations
in the coherent-state representation:
\begin{eqnarray}\label{ssd}
&&\bar{\alpha}\frac{\delta\rho(\bar{\alpha},\alpha^{\prime};t)}{\delta\bar{\alpha}}
\longleftrightarrow a^{\dag}a\rho(t),
\frac{\delta\rho(\bar{\alpha},\alpha^{\prime};t)}{\delta\alpha}\alpha \longleftrightarrow \rho (t)a^{\dag }a, \nonumber \\
&&~~~~~~~~~~~~\frac{\delta^{2}\rho(\bar{\alpha},\alpha
^{\prime};t)}{\delta\bar{\alpha}\delta\alpha} \longleftrightarrow
a\rho(t)a^{\dag},
\end{eqnarray}
with which we arrive at our final operator form of the non-Markovian
master equation
\begin{equation} \label{mas}
\dot{\rho}(t) = -\frac{i}{\hbar }[H^{\prime}(t),\rho (t)] +
\Gamma(t)[2a\rho(t)a^{\dag} - a^{\dag}a\rho(t) - \rho(t)a^{\dag}a],
\end{equation}
where $H^{\prime}(t) \equiv \hbar \Omega (t)a^{\dag }a$. This is the
exact master equation for the reduced system dynamics. Our exact
master equation (\ref{mas}) is similar to the non-Markovian master
equation of a two-level atom in a bosonic environment
\cite{Garraway97,Breuer02}. This similarity for the fully different
systems comes from the fact that both the derivations are based on
the rotating wave approximation and the initial vacuum state of the
environment. $\Omega(t)$, named as time-dependent Lamb shifted
frequency in the two-level atom system, plays here the role of a
time-dependent shifted frequency of the single-mode optical field
$S$ induced by the environment $E$. $\Gamma (t)$ represents a
time-dependent decay rate of the optical field.  We emphasize that
our derivation of the master equation goes beyond the Born-Markovian
approximation and contains all the backactions between $E$ and $S$
self-consistently. It can be seen that the only difference between
Eq. (\ref{mas}) and the master equation under Born-Markovian
approximation \cite{Carmichael93} is the time-dependent
coefficients. So we argue that all the non-Markovian characteristics
reside in the time-dependent coefficients of the generalized master
equation.

The time-dependent coefficients in our generalized master equation,
determined by Eq. (\ref{go}), essentially depend on the so-called
spectral density, which characterizes the coupling strength of the
environment to the system with respect to the environmental
frequencies.  It is defined as $J(\omega ) \equiv \sum_{k}\left\vert
g_{k}\right\vert^{2}\delta (\omega -\omega_{k})$.  In the continuous
limit of the environmental frequencies, we model in our work the
coupling of the optical field with the environment has a spectral
density as
\begin{equation}\label{sp}
J(\omega ) = \eta\omega\Big(\frac{\omega }{\omega
_{c}}\Big)^{n-1}e^{-\frac{\omega }{\omega _{c}}},
\end{equation}
where $\eta $ is a dimensionless coupling constant, and $\omega
_{c}$ is an exponential cutoff frequency.  The environment is
classified into three categories \cite{Leggett}: sub-Ohmic if
$0<n<1$, Ohmic if $n=1$, and super-Ohmic if $n>1$.  Different
spectral densities manifest different non-Markovian decoherence
dynamics \cite{Leggett}. The spectral density Eq. (\ref{sp}) is
motivated by the physical consideration that not all the modes of
the environment give the same contribution to the coupling to the
optical field system. Thus, the spectral density form is physically
reasonable to model the decoherence of our optical field system.
Actually such spectral density, especially the Ohmic one, is widely
used in the the decoherence analysis of optical fields in the
scenario of the continuous variable quantum information processing
\cite{Paris07,Goan07,An07,Horhammer08}.

Before presenting our numerical results in the next section, we show
how our generalized master equation reduces to the conventional
master equation by introducing the relevant Markovian approximation.
By defining new dynamical variables as $x(\tau) =
\alpha(\tau)e^{i\omega_0\tau}$, we can recast the first equation of
Eqs.(\ref{eom}) into
\begin{equation}\label{nn}
\dot{x}(\tau) + \int_0^{\infty}d\omega
J(\omega)\int_0^{\tau}d\tau^{\prime}e^{i(\omega_0 - \omega)(\tau -
\tau^{\prime})}x(\tau^{\prime}) = 0.
\end{equation}
Then, invoking the Markovian approximation, $x(\tau^{\prime}) \simeq
x(\tau )$, namely, approximately taking the dynamical variable to be
the one that depends only on the present time so that any memory
effect is ignored.  The Markovian approximation is mainly based on
the physical assumption that the correlation time of the environment
is much smaller compared with the typical time scale of the system
evolution.  Also under this assumption we can extend the upper limit
of the $\tau^{\prime}$ integration in Eq. (\ref{nn}) to infinity and
use the equality
\begin{equation}
\lim_{\tau \rightarrow \infty }\int_{0}^{\tau }d\tau ^{\prime
}e^{\pm i(\omega _{0}-\omega )(\tau -\tau ^{\prime })} = \pi \delta
(\omega -\omega_{0})\mp i\mathcal{P}\left( \frac{1}{\omega -\omega
_{0}}\right),
\end{equation}
where $\mathcal{P}$ denotes the Cauchy principal value.  The
integro-differential equation (\ref{nn}) thus reduces to a linear
ordinary differential equation.  The solution of $x(\tau )$, as well
as $\alpha (\tau )$ can then be obtained readily, which results in
\begin{equation}
u(\tau )=e^{-i(\omega _{0}-\delta \omega )\tau -\pi J(\omega
_{0})\tau },
\end{equation}
with $\delta \omega =\mathcal{P}\int_{0}^{\infty }\frac{J(\omega
)d\omega }{\omega -\omega _{0}}$.  Using this solution, one can
verify from Eq. (\ref{go}) that
\begin{equation}\label{mr}
\Gamma (t)=\pi J(\omega _{0}),~~\Omega (t)=\omega _{0}-\delta
\omega,
\end{equation}
which are precisely the coefficients in the Markovian master
equation of the optical system \cite{Carmichael93}.

In the next section, for definiteness, we consider Ohmic environment
$E$.  The characteristic time scale $\tau_E$ of the environmental
correlation function in the Ohmic case is roughly inversely
proportional to the cutoff frequency $\omega_c$ in Eq. (\ref{sp}),
i.e., $\tau _{E}\simeq 1/\omega _{c}$ \cite{Weiss}.  The cutoff
frequency $\omega _{c}$, which is originally introduced to eliminate
infinities in frequency integrations, therefore also determines if
the dynamics of open system $S$ is Markovian or non-Markovian.  Our
non-perturbatively derived exact results can allow us to explore all
these possibilities.

\section{Numerical results and discussions}
To illuminate the non-Markovian decoherence dynamics of $S$, we
consider the following initial state of the optical field:
\begin{equation}\label{scs}
\rho(0) = \frac{1}{N} [|\beta_0\rangle\langle\beta_0| +
|-\beta_0\rangle\langle -\beta_0| + |\beta_0\rangle\langle -\beta_0|
+ |-\beta_0\rangle\langle\beta_0|],
\end{equation}
where $N = 2(e^{\left\vert \beta _{0}\right\vert
^{2}}+e^{-\left\vert \beta_{0}\right\vert ^{2}})$ is a normalization
constant.  This is known as the Schr\"{o}dinger cat state and has
been produced experimentally \cite{Sch-kitten, Sch-cat}.  After some
straightforward calculations, we obtain, via Eqs. (\ref{rout}) and
(\ref{prord}),
\begin{eqnarray}\label{tt}
\rho(t) &= &\frac{1}{N} [e^{\left\vert\beta_0\right\vert^2 -
\left\vert\beta\right\vert^2}(|\beta\rangle\langle\bar{\beta}| +
|-\beta\rangle\langle -\bar{\beta}|) \nonumber \\&&+
 e^{-(\left\vert\beta_0\right\vert^2 - \left\vert\beta\right\vert^2)}(|-\beta\rangle\langle\bar{\beta}| + |\beta\rangle\langle -\bar{\beta}|)],
\end{eqnarray}
where $\beta = \beta_0u(t)$.  From Eq. (\ref{tt}), the purity which
is defined as $p(t) = Tr\rho^2(t)$ can be calculated readily as
\begin{eqnarray}
p(t) &=&\frac{2}{N^{2}}[e^{2\left\vert \beta
_{0}\right\vert^{2}}+e^{-2\left\vert \beta _{0}\right\vert
^{2}}+e^{2\left\vert \beta _{0}\right\vert ^{2}-4\left\vert \beta
\right\vert^{2}}\nonumber \\&& +e^{-2\left\vert \beta
_{0}\right\vert ^{2} +4\left\vert \beta\right\vert ^{2}}+4].
\end{eqnarray}

In Fig. \ref{fr1}, we plot the numerical results of the decay rate
$\Gamma(t)$, frequency shift $\delta\omega(t)$, and purity $p(t)$
when $\tau_E \ll \tau_0$.  We note that the exact $\Gamma(t)$ and
$\delta\omega(t)$ differ from their corresponding values obtained
via a Markovian treatment only over a very short time interval.  For
$t < \tau_E$, both coefficients grow very quickly, while for $t >
\tau_E$, they gradually approach the corresponding Markovian
results, given by Eqs. (\ref{mr}), as $t$ approaches $\tau_0$.  The
finite, almost steady, positive decay rate guarantees the
irreversibility of the system dynamics.  Accordingly, the exact time
evolution of the purity shows only slight deviation from the
Markovian results.  The Schr\"{o}dinger cat state eventually evolves
to a steady state, namely the ground state of $S$: $\rho_g =
|0\rangle\langle0|$.  Clearly, in this case, the backaction of the
environment $E$ has negligible effect on the dynamics of system $S$,
and the Markovian approximation is applicable.  We say the system
dynamics is mainly governed by the dissipative effect of $E$.

Figure \ref{fre} shows the decay rate $\Gamma(t)$, frequency shift
$\delta\omega(t)$, and purity $p(t)$, when $\tau_E = \tau_0$.  In
this case, the backaction of the environment $E$ has a considerable
impact on the dynamics of our system $S$, and the Markovian
approximation is not applicable. Firstly, we note that, in contrast
to the Markovian treatment, the decay rate can take negative values.
Physically, this corresponds to $S$ reabsorbing a photon from $E$,
which will lead to an increase in the photon number of $S$
\cite{Breuer02}.  Next, more interestingly, we note that the decay
rate approaches zero asymptotically, which dramatically differs from
the Markovian results.  Consequently, the exact evolution of the
Schr\"{o}dinger cat state is drastically different from the
Markovian results.  In particular, it eventually evolves to some
steady state, which is not the ground state $\rho_g$.  The
backaction of the environment causes the system to undergo transient
oscillations, which is characteristic of non-Markovian dynamics.
From previous studies \cite{Ban06, Goan07, An07}, one would have
concluded that non-Markovian effects only show up in short-time
dynamics.  Our results, however, clearly show on the contrary that
non-Markovian effects can also have an influence on the long-time
behavior of the system dynamics and the final steady state of $S$.
This is because that the dissipative effect of $E$ is strongly
counteracted by the effect due to the backaction of $E$.

The results under the condition $\tau_E \gg \tau_0$ is shown in Fig.
\ref{fre3}.  Because of the extremely long memory effect of the
environment, the backaction from $E$ to $S$ is so strong that it
governs the dynamics of $S$.  This causes the decay rate to
oscillate over a very long duration.  This is typical of
non-Markovian dynamics \cite{Paris07}.  The `equilibrium' position
for the oscillation of $\Gamma(t)$ is not at zero, but a small
positive value.  The positivity of the equilibrium position of the
decay rate oscillation makes the system dynamics experiences weak
dissipation.  Such weak dissipation is verified by the time
evolution of the purity of the Schr\"{o}dinger cat state in Fig.
\ref{fre3}. The evolution of $p(t)$ also shows that the coherent
oscillation induced by the backaction of $E$ persists for a very
long time, even as the state approaches the ground state.

In summary, depending on $\tau_E$ in comparison to $\tau_0$, the
decoherence dynamics of $S$ shows different behaviors.  For $\tau_E
\ll \tau_0$, the exact results show only slight deviation from the
Markovian ones.  Since the effect due to backaction of $E$ is
extremely small, the system dynamics is mainly governed by the
dissipative effect of $E$, and the widely used Markovian
approximation is applicable.  For $\tau_E = \tau_0$, the
considerable backaction induced by the near resonant interaction
between $E$ and $S$ counteracts the dissipative effect, and results
in zero decay rate asymptotically.  This causes dissipation to cease
before the system decays to its ground state.  That is, the steady
state is no longer the ground state, unlike the previous case.  When
$\tau_E \gg \tau_0$, the backaction of $E$ is so strong that it
governs the dynamics of $S$.  The decay rate of the system
oscillates about some equilibrium position over a very long period
of time.  The positivity of such an equilibrium position guarantees
the overall weak dissipation effect of $E$.

\section{Conclusion}
In this work, using the influence-functional method of Feynman and
Vernon, we investigate the exact decoherence dynamics of a
single-mode optical field $S$ in an environment at zero temperature.
We derive an exact generalized master equation for $S$.  The
equation's time-dependent coefficients depend on the environmental
spectral density.  These determine the exact dynamics of $S$.  We
conclude from our numerical results that $E$ exerts two competing
influences on $S$. One effect, ${\cal E}_1$, is dissipation, and is
responsible for the decoherence of $S$.  The other, ${\cal E}_2$, is
due to the backactions of $E$ on $S$.  The backaction of $E$ on $S$
means that $E$ with its state changed due to interaction with $S$,
in turn, exerts its dynamical influence back on $S$. This is the
physical origin of the non-Markovian dynamics of $S$.  In the
conventional Born-Markovian treatment, one generally neglects the
backaction of the environment.  So, in that case, we do not see the
effect due to backaction on the dynamics of $S$.  The degree of
manifestation of ${\cal E}_1$ and ${\cal E}_2$ in the dynamics of
$S$ depends on $\tau_E$ in comparison with $\tau_0$.  For $\tau_E
\ll \tau_0$, ${\cal E}_1$ dominates and ${\cal E}_2$ only gives rise
to a transient coherent oscillation of $S$. The state of $S$ evolves
to the ground state $\rho_g$, which is coincident with the Markovian
result.  If $\tau_E= \tau_0$, ${\cal E}_1$ and ${\cal E}_2$ are
comparable and their effects counteract each other. The state of $S$
asymptotically evolves to some steady state, which is not the ground
state of $S$. Finally, when $\tau_E \gg \tau_0$, ${\cal E}_2$
dominates and governs the dynamics of $S$.  The state of $S$
eventually approaches to the ground state but never quite reach it.

The theory we have established is a non-perturbative description of
the exact decoherence dynamics of a single-mode optical field.  It
is a generalization of the well-developed Born-Markovian treatment
of such system.  It can serve as a useful basic theoretical model in
analyzing the non-Markovian decoherence dynamics of optical fields
employed in practical quantum information processing schemes.  It
should be noted that although only the Ohmic spectral density is
considered here, it is straightforward to generalize our discussion
to the non-Ohmic cases.

\section*{Acknowledgement}
The work is supported by NUS Research Grant No. R-144-000-189-305.
J.H.A. also thanks the financial support of the NNSF of China under
Grant No 10604025, and the Fundamental Research Fund for Physics and
Mathematics of Lanzhou University under Grant No Lzu05-02.

\end{document}